\documentclass[10pt,conference]{IEEEtran}

\usepackage{cite}
\usepackage{amsmath}
\usepackage{amssymb}
\usepackage{graphicx}
\usepackage{booktabs}
\usepackage[caption=false]{subfig}
\usepackage{minted}
\usepackage{etoolbox}
\usepackage{mathpartir}
\usepackage{tikz}
\usetikzlibrary{shapes.geometric}
\usepackage{hyperref}
\usepackage{fancyvrb}

\setminted{fontsize=\scriptsize}
\BeforeBeginEnvironment{minted}{\vspace{4pt}}
\AfterEndEnvironment{minted}{\vspace{4pt}}

\newcommand{\ket}[1]{|#1\rangle}

\DefineVerbatimEnvironment{pseudocode}{Verbatim}{
  commandchars=\\\{\},
      fontsize=\scriptsize
}

\title{qstack: Compositional End-to-End Compilation for Fault-Tolerant Quantum Programs}

\author{
\IEEEauthorblockN{Andr\'es Paz and Dan Grossman}
\IEEEauthorblockA{\textit{University of Washington} \\
Seattle, WA, USA \\
\{anpaz, djg\}@cs.washington.edu}
}

\begin{document}

\maketitle

\begin{abstract}
Compiling quantum programs for fault-tolerant execution requires transforming high-level operations through multiple abstraction layers: from logical gates to error-corrected encodings to hardware-native instructions. A key challenge is that quantum error correction turns purely quantum programs into hybrid quantum-classical programs, where classical feedback from syndrome measurements drives quantum corrections at runtime. Existing compilation frameworks handle these quantum and classical components separately, requiring manual adaptation of classical logic at each compilation stage, all while preserving program semantics.

We present \textbf{qstack}, a compiler framework built around a purely quantum intermediate representation in which classical logic is accessed only through opaque callbacks, written in any classical language. The framework's central mechanism, \emph{callback wrapping}, enables compositional compilation: each compiler pass automatically adapts both quantum operations and their associated classical callbacks, and any kernel dynamically generated by a callback is compiled through the full pipeline. This allows ISA translation and quantum error correction to be expressed as composable compiler passes, including concatenation of error-correcting codes, without manual intervention. We demonstrate end-to-end compilation from a high-level gate set through Clifford gates to trapped-ion native operations, with bit-flip and phase-flip repetition codes, the Steane code, and the Shor code obtained by composing two repetition passes.
\end{abstract}

\section{Introduction}\label{sec:introduction}

Quantum programs are inherently hybrid. While it is tempting to think of them as simple lists of quantum instructions with measurements at the end, practical algorithms require classical control interleaved with quantum operations. Teleportation needs classical fixups based on measurement outcomes, T-gate injection requires conditional corrections, and quantum error correction (QEC) demands repeated cycles of syndrome measurement, classical decoding, and quantum correction. In each case, intermediate measurements produce classical data that determines what quantum operations come next.

\subsection{Motivation}

This hybrid nature compounds at compilation time. A quantum program written against a high-level gate set must be compiled to a canonical gate set and then to hardware-native operations. If higher fidelity is needed, one or more error correction layers are inserted along the way. Each QEC layer converts a single logical operation into a cycle of quantum gates, syndrome measurements, classical decoding, and conditional corrections—introducing its own layer of classical feedback on top of any the algorithm already contains. The result is multiple layers of classical feedback that must compose correctly across the entire pipeline.

Whether frameworks embed circuit construction in a host language~\cite{Qiskit, cirq, green2013quipper} or provide native classical control alongside quantum operations~\cite{cross2022openqasm, smith2016practical, qsharp, silq, guppy}, the compiler must ultimately deal with the interaction between multiple layers of classical feedback. When a source program contains classical control (e.g., a repeat-until-success loop) and is then compiled through error correction, the compiler must handle both the algorithm's own measurement-driven logic and the error correction's syndrome-driven logic. These layers must compose correctly, and any quantum instructions generated dynamically by the algorithm's classical logic must themselves be compiled through the full pipeline.

In current frameworks, this composition is manual. When a QEC compiler expands a logical qubit into physical qubits, the developer must adapt the classical decoding logic for each compilation stage and each level of code concatenation~\cite{paetznick2024demonstration,reichardt2024tesseract,reichardt2024faulttolerant}. Alternatively, one can express decoders using the circuit's own classical primitives, which makes composition possible but requires each compiler pass to correctly recurse into all control-flow blocks and remap classical registers across layers. In Section~\ref{sec:evaluation} we examine both approaches in detail. A more detailed comparison with existing frameworks appears in Section~\ref{sec:related}.

\subsection{Our Approach: Compositional Compilation via Callback Wrapping}

We present \textbf{qstack}, a compiler framework built around two ideas. First, a \emph{uniform IR} in which quantum programs are represented as kernels with opaque classical callbacks: the IR captures quantum instructions and measurement outcomes, while all classical logic lives externally, written in any language (Section~\ref{sec:ir}). Second, \emph{callback wrapping}: when a compiler pass transforms a program, it automatically adapts each callback so that it receives correctly decoded measurement results and any kernel it generates is compiled through the full pipeline (Section~\ref{sec:compilation}). Together, these yield several concrete properties:

\begin{itemize}
    \item \textbf{Uniform representation.} The same IR format describes programs at every stage of the compilation pipeline. Because the IR is purely quantum and the classical side is unconstrained, the compiler transforms gates, rewrites qubit allocations, and inserts error correction without reasoning about classical control flow.

    \item \textbf{ISA translation and QEC are both compiler passes.} The same mechanism handles instruction set translation and error correction.
    
    \item \textbf{Passes compose freely.} Passes chain whenever the output gate set of one matches the input of the next. ISA translation passes change the instruction set, while QEC passes preserve it (e.g., Cliffords $\to$ Cliffords), so QEC layers can be inserted or concatenated at any point.
    
    \item \textbf{Classical logic adapts automatically.} A callback written once works unchanged through all compilation stages; the adapter chain ensures it receives decoded measurements and that any kernels it produces are fully compiled. No compiler pass needs to know about any other pass in the chain.
    
    \item \textbf{Mid-circuit correction.} Callbacks execute during evaluation, so syndrome-driven corrections are applied between quantum operations.
\end{itemize}

\subsection{Contributions}

We make the following contributions:

\begin{enumerate}
    \item A \textbf{compiler framework} for quantum programs in which both ISA translation and quantum error correction are expressed as composable compiler passes that automatically adapt classical callbacks through a wrapping mechanism (Section~\ref{sec:compilation}).
    
    \item A \textbf{minimal IR} based on a kernel model with opaque classical callbacks, providing a uniform representation across all compilation stages with fully formalized syntax and semantics (Sections~\ref{sec:ir} and~\ref{sec:formal}).
    
    \item A \textbf{demonstration and evaluation} of end-to-end compilation from a high-level gate set through Clifford gates to trapped-ion native operations, with repetition-code and Steane code QEC layers that compose freely, including construction of the Shor code by composing bit-flip and phase-flip repetition passes. We compare the composability and structure of qstack programs against equivalent Qiskit implementations (Section~\ref{sec:evaluation}).
\end{enumerate}

Our implementation is a Python framework of approximately 1,500 lines, including a state-vector emulator, noise models, and Jupyter integration. The source code and examples are available as supplementary material.

\section{The qstack Intermediate Representation}\label{sec:ir}

The representation centers on two constructs: \emph{kernels}, which structure quantum computation around qubit allocation, instruction, and measurement; and \emph{gate sets}, which specify the available operations. A \textbf{program} pairs a kernel with a gate set.

\subsection{Kernels}

The fundamental unit of a qstack program is the \emph{kernel}. A kernel follows a fixed four-phase cycle:

\begin{minipage}{\linewidth}
\begin{Verbatim}[commandchars=\\\{\},fontsize=\scriptsize]
allocate \emph{q}:
  \emph{instructions}
measure \emph{callback}
\end{Verbatim}
\end{minipage}

\noindent\texttt{allocate} introduces a fresh qubit \emph{q}. The body contains a sequence of \emph{instructions}, each of which is either a quantum gate applied to one or more qubits, or a nested kernel. When the body completes, the qubit is measured and deallocated; the measurement outcome (a classical bit) is passed to a \emph{callback}.

Kernels nest: a kernel may appear anywhere an instruction is expected. A nested kernel can apply gates to any qubit allocated by an enclosing kernel, so instructions have access to all qubits currently in scope. Because each kernel allocates exactly one qubit and measures it at the end, qubits follow a strict stack discipline (LIFO allocation and deallocation). This guarantees that every qubit has a correct lifecycle: it is initialized, used, and measured exactly once.

\subsection{Gate Sets}

A gate set is a named collection of quantum gate definitions. Each gate definition specifies a name, the number of target qubits, and its unitary matrix (or a parameterized factory that produces one) which can be used for simulation and validation. Figure~\ref{fig:gate-set} shows how gates for a gate set can be defined.

\begin{figure}[t]
\begin{Verbatim}[fontsize=\tiny]
r2 = 1/sqrt(2)

X  = QuantumDefinition("x",  targets=1, matrix=[[0,1],[1,0]])
H  = QuantumDefinition("h",  targets=1, matrix=[[r2,r2],[r2,-r2]])
CX = QuantumDefinition("cx", targets=2, matrix=[[1,0,0,0],[0,1,0,0],
                                                [0,0,0,1],[0,0,1,0]])

# Parameterized gates use a factory that produces a matrix from arguments:
U1 = QuantumDefinition.with_parameters("u1", targets=1, factory=u1_matrix)
RZ = QuantumDefinition.with_parameters("rz", targets=1, factory=rz_matrix)
\end{Verbatim}
\caption{Gate definitions. Fixed gates specify a unitary matrix directly; parameterized gates (e.g., those used in the native set for a trapped-ion device) use a factory method that computes the matrix from gate parameters.}
\label{fig:gate-set}
\end{figure}

The framework ships with four gate sets used in our evaluation: a pedagogical ``Toy'' set (\texttt{flip}, \texttt{mix}, \texttt{entangle}), a Clifford set ($X$, $Y$, $Z$, $S$, $H$, $CX$, $CZ$), a trapped-ion native set~\cite{moses2023race} ($U_1$, $R_Z$, $ZZ$), and a neutral-atom native set~\cite{bluvstein2024logical} ($RZ$, $CZ$, $SX$). Users can define additional sets. A compiler pass declares a \emph{source} and \emph{target} gate set; passes chain when one's target matches the next's source.

\subsection{Callbacks}

A callback is an opaque classical function that receives a classical bit list and optionally returns a new kernel. The IR stores only the callback's name; the implementation lives externally, in Python or any host language. If a callback returns a kernel, that kernel is executed immediately; if it returns \texttt{None}, execution of the current kernel terminates and control returns to the enclosing kernel.

The default callback, \texttt{done}, is a no-op: leaves the classical bit list intact and returns no new kernel, ending execution of the current kernel. More complex callbacks pop bits from the state to implement error correction decoders, repeat-until-success logic, or conditional fixups. 

Callbacks may also accept one or more \emph{parameters}, which are constant values stored in the IR (for example, a qubit identifier to apply a correction to). These parameters are fixed at program construction time and visible to the compiler, while the callback body remains opaque.

Because callbacks are opaque, the IR contains no classical expressions, no classical variables, and no classical control flow. The compiler sees only quantum instructions and callback references. All classical computation is deferred to the callback boundary, where it is handled by the wrapping mechanism described in Section~\ref{sec:compilation}.

Importantly, the goal of this work is not to design a classical language for callbacks. The IR is purely quantum; callbacks can be written in any classical language, with arbitrary control flow, data structures, or external library calls.

There may be practical reasons to restrict the expressivity available to callbacks (for example, to guarantee termination, bound execution time, or enable classical compilation to a specific control processor). Such restrictions are a valid concern, but one that can be enforced within this framework: the callback interface enables a natural point at which to impose a restricted classical language or a static analysis pass, without changing the quantum IR or the compilation pipeline. Similarly, classical optimizations (such as branch elimination or constant folding) are not lost; they are simply delegated to the classical language's own compiler, which can apply them independently. Because the quantum IR and the classical language are independent, the quantum compiler is free to transform programs without reasoning about classical semantics, and the classical language is free to evolve (or be constrained) without affecting the quantum compilation pipeline.

\subsection{Evaluation}

Evaluation proceeds depth-first. Starting a kernel extends the quantum state by one qubit (tensoring with $\ket{0}$) and maps the qubit identifier to it. Gates apply their unitaries to the quantum state; nested kernels are evaluated recursively. When a kernel's body completes, its qubit is measured in the computational basis, yielding a classical bit that is pushed onto a measurement stack, and its callback is invoked. The callback may return a new kernel, which is then evaluated immediately; otherwise control returns to the enclosing kernel.

Because each kernel allocates one qubit on entry and measures it on exit, evaluation gives rise to two stacks. The \emph{kernel stack} tracks active invocations: starting a kernel pushes it; measuring its qubit pops it. The \emph{measurement stack} accumulates classical outcomes: each measurement pushes one bit. Callbacks receive the full measurement stack and may consume bits via \texttt{pop} or produce bits via \texttt{push}. The two stacks evolve in lockstep: when a kernel completes, it leaves the kernel stack and its measurement outcome enters the measurement stack. When the program terminates, the measurement stack contains the program's classical output. The full evaluation rules are formalized in Section~\ref{sec:formal}.

\subsection{Example: Repeat-Until-Success}

The following program illustrates the IR and its evaluation:

\begin{minipage}{\linewidth}
\begin{Verbatim}[fontsize=\scriptsize]
allocate q1:
  allocate q2:
    h q2
    cx q2 q1
  measure done
measure repeat_until_zero
\end{Verbatim}
\end{minipage}

\noindent The inner kernel allocates \texttt{q2}, applies a Hadamard and a CNOT to create a Bell state, and measures \texttt{q2} with the default \texttt{done} callback. The outer kernel then measures \texttt{q1} with a \texttt{repeat\_until\_zero} callback, defined in the host language (in this case Python) as a regular function:

\begin{Verbatim}[fontsize=\scriptsize]
def bell_kernel():
    return Kernel("q1", [
        Kernel("q2", [
          H("q2"), CX("q2", "q1")
        ], done)
    ], repeat_until_zero)

def repeat_until_zero(mstack):
    if mstack.pop() == 1:
        return bell_kernel()
    return None
\end{Verbatim}

\noindent If the measurement result is 1, the callback returns a new kernel identical to the original, and the process repeats. If the result is 0, the callback returns \texttt{None} and the program terminates. This is a complete, executable qstack program: no classical control flow appears in the IR, yet the program implements an unbounded probabilistic loop.

\begin{figure}[t]
\centering
\includegraphics[width=\columnwidth]{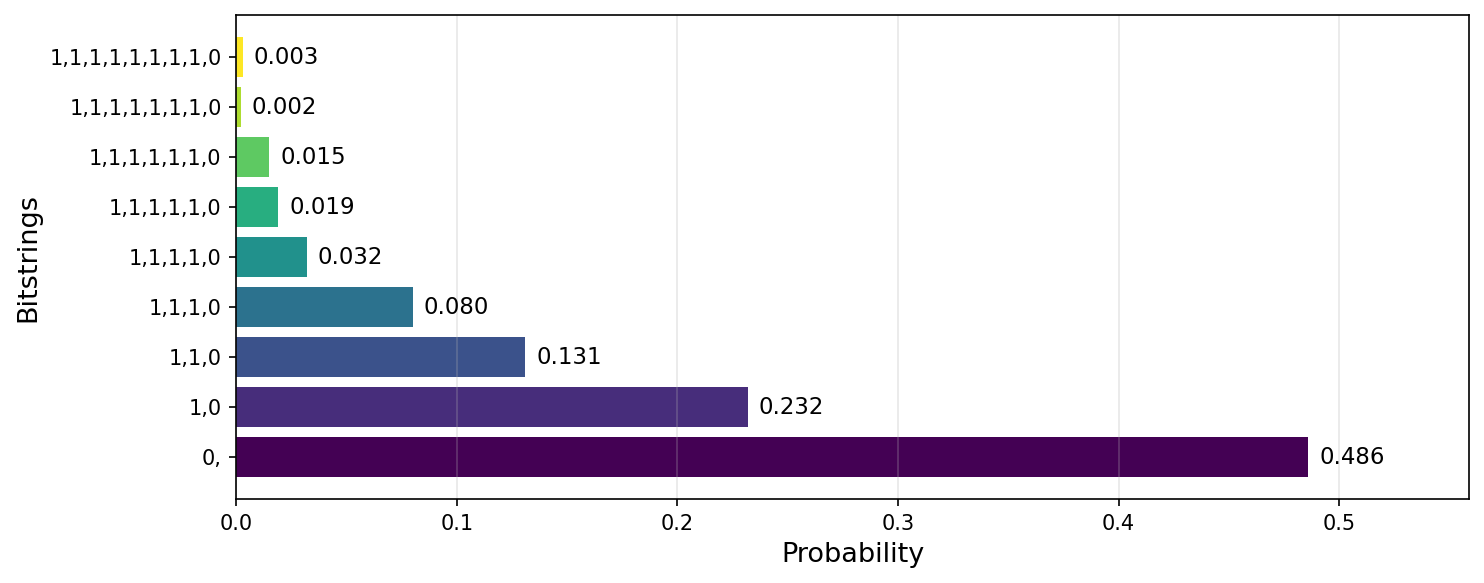}
\caption{Distribution of measurement outcomes from 1000 simulation runs of the repeat-until-success program. Bitstrings of length $n$ occur with probability $1/2^{n}$, confirming the expected geometric decay.}
\label{fig:repeat-until-success}
\end{figure}

When the program terminates, the measurement stack contains a~0 (the final successful measurement) followed by zero or more~1s, one for each iteration in which the callback observed an unsuccessful outcome and restarted the kernel. Because each measurement outcome is equally likely, the probability of observing exactly $k$ trailing 1s is $1/2^{k+1}$, halving with each additional repetition. Figure~\ref{fig:repeat-until-success} confirms this: a histogram of 1000 simulation runs shows the expected geometric decay, and the observed frequencies closely match the theoretical probabilities.

Tracing this example illustrates the role of the kernel stack and measurement stack introduced above. Starting the outer kernel pushes it onto the kernel stack; starting the inner kernel pushes it on top. When \texttt{q2} is measured, the inner kernel is popped and its bit is pushed onto the measurement stack; the \texttt{done} callback leaves both stacks unchanged. When \texttt{q1} is measured, the outer kernel is popped and its bit is pushed. The \texttt{repeat\_until\_zero} callback pops one bit from the measurement stack; if it is~1, a new kernel is pushed onto the kernel stack and the process repeats.

\section{Compositional Compilation via Callback Wrapping}\label{sec:compilation}

Because the qstack IR provides a uniform representation at every stage of the compilation pipeline, a compiler pass is always a function from kernel to kernel: it consumes a kernel in one gate set and produces a semantically equivalent kernel in another (or the same) gate set. Passes serve many purposes: translating between gate sets, injecting error correction, optimizing gate counts, or reordering instructions among others.

\subsection{Gate Compilation via Handlers}

The framework structures gate compilation around \emph{handlers}: a compiler pass provides one handler per source gate, and the framework applies them while traversing the kernel AST.

For a simple ISA translation pass, each handler maps a single source gate to one or more target gates. For example, the Toy-to-Cliffords compiler defines three handlers:

\begin{Verbatim}[fontsize=\scriptsize]
def handle_flip(inst):       return X(inst.targets[0])
def handle_mix(inst):        return H(inst.targets[0])
def handle_entangle(inst):   return CX(inst.targets[0],
                                       inst.targets[1])
\end{Verbatim}

\noindent Each handler receives a gate from the source gate set and returns one or more gates in the target set. The framework traverses the kernel recursively: when it encounters a gate, it invokes the corresponding handler; when it encounters a nested kernel, it recurses. The compiler author writes only the per-gate logic.

For QEC passes the handlers are more involved. The framework includes two repetition-3 codes: a \emph{bit-flip} code (rep-3 bit) that protects against $X$ errors, and a \emph{phase-flip} code (rep-3 phase) that protects against $Z$ errors. In the bit-flip code, a handler for $X$ returns a \emph{kernel} that applies $X$ to each of the three physical qubits encoding the logical qubit. This is a key feature of the framework: because kernels and gates are both valid instructions in the IR (Section~\ref{sec:ir}), a handler can return either one. A QEC handler can therefore expand a single logical gate into a multi-qubit sub-kernel that allocates ancillae, applies physical gates, and performs syndrome measurements, all within the same IR.

The framework does not prevent more advanced transformations; at any point the compiler has access to the entire kernel and can perform global optimizations such as gate cancellation or qubit routing.

\subsection{Callback Wrapping}

When a compiler pass changes the measurement structure (for example, a QEC pass that encodes one logical qubit into three physical qubits, producing three measurement bits where there was one), the classical bit list no longer matches what the original callback expects. The callback was written to pop one bit; after compilation it will find three. To preserve semantics, the compiler must restore the bit list to the form the callback was written for.

The qstack solution is to \emph{wrap} each callback in an adapter that restores stack consistency. The wrapper interposes a \textbf{decoder} that pops the $n$ physical bits from the measurement stack and pushes back the decoded logical bit. After the decoder runs, the measurement stack looks exactly as the original callback expects, and the callback runs unchanged.

Since the only other action a callback can take is to return a new kernel, and since that kernel is expressed in the \emph{source} instruction set, the wrapper also \textbf{compiles} any returned kernel by passing it back through the same compiler. Since the decoder is the only part that varies, the framework factors it out. The wrapping boilerplate is provided by the framework; a concrete compiler only needs to implement a \texttt{decode} method that restores the classical stack. In pseudocode:

\begin{Verbatim}[fontsize=\scriptsize]
# Framework boilerplate (same for every compiler pass):
def wrapped_callback(mstack):
    compiler.decode(mstack)             # restore the stack
    kernel = original_callback(mstack)  # invoke original
    if kernel is not None:
        return compiler.compile_kernel(kernel)  # compile
    return None

# Bit-flip rep-3 compiler provides only this:
def decode(self, mstack):
    m0 = mstack.pop()
    m1 = mstack.pop()
    m2 = mstack.pop()
    mstack.push(majority_vote(m0, m1, m2))
\end{Verbatim}

For a simple ISA translation pass (e.g., Toy $\to$ Cliffords), the decode method is a no-op because the measurement structure is unchanged. For a QEC pass, it implements the code's classical decoding algorithm: majority vote for the repetition codes, syndrome decoding for Steane.

Each decoder only touches the top of the measurement stack. A QEC pass replaces one logical qubit with $n$ physical qubits; the LIFO discipline guarantees that the $n$ physical measurement bits sit on top of the measurement stack. The decoder pops exactly those $n$ bits and pushes back one decoded logical bit; the rest of the stack is untouched. Moreover, inner kernels complete before outer kernels, so their wrappers fire first: by the time an outer callback runs, any bits placed by inner kernels are already decoded to logical bits. Although the callback has access to the full stack, correctness follows from the fact that each decoder only needs to process its own encoding's bits, and inner decoders have already restored their portion of the stack.

\paragraph*{Example} Consider the repeat-until-success program from Section~\ref{sec:ir} compiled through the bit-flip rep-3 pass. Each logical qubit is encoded into three physical qubits, and the compiled program becomes:

\begin{Verbatim}[fontsize=\scriptsize]
allocate q1.0:
 allocate q1.1:
  allocate q1.2:
   allocate q2.0:
    allocate q2.1:
     allocate q2.2:
      h q2.0; h q2.1; h q2.2
      cx q2.0 q1.0; cx q2.1 q1.1; cx q2.2 q1.2
     measure done          # q2.2 (new physical qubit)
    measure done           # q2.1 (new physical qubit)
   measure w_done          # q2.0 (original done, wrapped)
  measure done             # q1.2 (new physical qubit)
 measure done              # q1.1 (new physical qubit)
measure w_repeat           # q1.0 (original repeat, wrapped)
\end{Verbatim}

\noindent The newly introduced physical qubits use plain \texttt{done} callbacks. The original callbacks are wrapped: \texttt{w\_done} on \texttt{q2.0} pops three physical bits, majority-decodes them into one logical bit, and pushes it back; \texttt{w\_repeat} does the same for \texttt{q1}'s three physical bits before invoking the original \texttt{repeat\_until\_zero}. 

The following table traces the measurement stack through both programs. Every time the original callback is invoked, the compiled program's measurement stack matches the original exactly. If a wrapped callback returns a new kernel, the wrapper compiles it through the same rep-3 pass before returning it for execution.

\medskip
\noindent
\scriptsize
\setlength{\tabcolsep}{3pt}
\begin{tabular}{@{}lll|lll@{}}
\toprule
\multicolumn{3}{c|}{\textbf{Original}} & \multicolumn{3}{c}{\textbf{Compiled (rep-3 bit)}} \\
\textbf{event} & \textbf{action} & \textbf{mstack} & \textbf{event} & \textbf{action} & \textbf{mstack} \\
\midrule
& & & measure q2.2 & & $m_{22}$ \\
& & & \texttt{done} & --- & $m_{22}$ \\[2pt]
& & & measure q2.1 & & $m_{21} :: m_{22}$ \\
& & & \texttt{done} & --- & $m_{21} :: m_{22}$ \\[2pt]
measure q2 & & $m_2$ & measure q2.0 & & $m_{20} :: m_{21} :: m_{22}$ \\
& & & \texttt{w\_done} & decode & $\boldsymbol{m_2}$ \\
\texttt{done} & --- & $m_2$ & \texttt{done}  & & $m_2$ \\[2pt]
& & & measure q1.2 & & $m_{12} :: m_2$ \\
& & & \texttt{done} & --- & $m_{12} :: m_2$ \\[2pt]
& & & measure q1.1 & & $m_{11} :: m_{12} :: m_2$ \\
& & & \texttt{done} & --- & $m_{11} :: m_{12} :: m_2$ \\[2pt]
measure q1 & & $m_1 :: m_2$ & measure q1.0 & & $m_{10} :: m_{11} :: m_{12} :: m_2$ \\
& & & \texttt{w\_repeat} & decode & $\boldsymbol{m_1 :: m_2}$ \\
\texttt{repeat} & pop & $m_2$ & \texttt{repeat}  & pop & $m_2$ \\[2pt]
\bottomrule
\end{tabular}
\normalsize
\medskip

\noindent The \texttt{decode} step pops three physical bits and pushes one logical bit (majority vote), restoring the mstack to the form the original callback expects. After decoding (bold), the compiled mstack matches the original.

\subsection{Composability}\label{sec:composability}

Callback wrapping composes across compiler passes. Consider two passes applied in sequence: the program is first compiled by pass 1, then by pass 2. Each pass wraps the callbacks it receives. After pass 1, each original callback is wrapped with pass 1's decoder and compiler. After pass 2, that already-wrapped callback is wrapped again with pass 2's decoder and compiler. The result is a chain of adapters, one per pass.

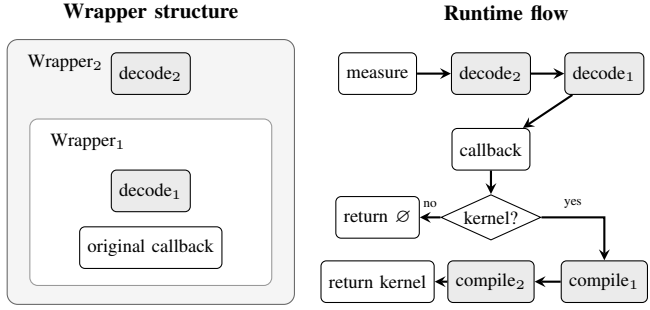
\begin{figure}[t]
\centering
\begin{tikzpicture}[
  box/.style={rectangle, draw, rounded corners=2pt, minimum height=0.55cm,
              font=\scriptsize, inner sep=3pt},
  arr/.style={->, thick, >=stealth},
  darr/.style={->, thick, >=stealth, dashed},
  lbl/.style={font=\scriptsize, midway},
]
  \node[font=\footnotesize\bfseries] at (1.8, 3.55) {Wrapper structure};

  \draw[rounded corners=4pt, draw=black!60, fill=black!4]
    (-0.1, -0.3) rectangle (3.7, 3.2);
  \node[font=\scriptsize, anchor=north west] at (0.05, 3.12) {Wrapper$_2$};

  \draw[rounded corners=3pt, draw=black!40, fill=white]
    (0.2, -0.05) rectangle (3.4, 2.15);
  \node[font=\scriptsize, anchor=north west] at (0.35, 2.08) {Wrapper$_1$};

  \node[box, fill=white] (cb) at (1.8, 0.45) {original callback};

  \node[box, fill=black!8] (d2) at (1.8, 2.75) {decode$_2$};
  \node[box, fill=black!8] (d1) at (1.8, 1.20) {decode$_1$};

  \node[font=\footnotesize\bfseries] at (6.5, 3.55) {Runtime flow};

  \node[box] (meas) at (4.8, 2.75) {measure};
  \node[box, fill=black!8] (rd2) at (6.3, 2.75) {decode$_2$};
  \node[box, fill=black!8] (rd1) at (7.8, 2.75) {decode$_1$};

  \node[box] (rcb) at (6.3, 1.75) {callback};

  \node[diamond, draw, inner sep=1pt, font=\scriptsize, aspect=2.2] (kq) at (6.3, 0.85) {kernel?};

  \node[box, fill=black!8] (rc1) at (7.8, 0.0) {compile$_1$};
  \node[box, fill=black!8] (rc2) at (6.3, 0.0) {compile$_2$};

  \node[box] (ret) at (4.8, 0.0) {return kernel};
  \node[box] (rn)  at (4.8, 0.85) {return $\varnothing$};

  \draw[arr] (meas) -- (rd2);
  \draw[arr] (rd2) -- (rd1);
  \draw[arr] (rd1) -- (rcb);

  \draw[arr] (rcb) -- (kq);

  \draw[arr] (kq) -| node[lbl, above, pos=0.25] {\tiny yes} (rc1);
  \draw[arr] (rc1) -- (rc2);
  \draw[arr] (rc2) -- (ret);

  \draw[arr] (kq) -- node[lbl, above] {\tiny no} (rn);

\end{tikzpicture}
\caption{Composability of two compiler passes. \textbf{Left}: the wrapping structure after pass~1 and pass~2 each wrap the original callback. \textbf{Right}: runtime data flow when a measurement triggers the outermost wrapper. Decoders run outside-in (pass~2 then pass~1); if the callback returns a kernel, compilers run inside-out (pass~1 then pass~2).}
\label{fig:composability}
\end{figure}

At runtime (Figure~\ref{fig:composability}), when the outermost wrapper is invoked after a measurement, it decodes pass 2's measurements (restoring the stack) and calls the inner wrapper, which decodes pass 1's measurements and calls the original callback. If the original callback returns a new kernel (expressed in the source gate set), the inner wrapper compiles it through pass 1, and the outer wrapper compiles the result through pass 2. The dynamically generated kernel goes through the same compilation pipeline as the original program, in the same order. Each wrapper knows only about its own compiler; no pass needs awareness of any other pass in the chain. This extends to any number of passes: $n$ passes produce $n$ layers of wrapping, and every dynamically generated kernel is compiled through all $n$ passes in order.

This composability applies directly to QEC. Since a QEC pass preserves the gate set (e.g., Cliffords $\to$ Cliffords), passes can be composed freely. Composing the bit-flip rep-3 pass with the phase-flip rep-3 pass yields the Shor code (9 physical qubits per logical qubit), a full quantum error-correcting code, even though neither repetition code alone corrects arbitrary errors. Similarly, composing a Steane pass with a rep-3 pass yields a concatenated Steane-over-rep-3 code, and applying the same pass twice yields a distance-doubled concatenated code. Each composition adds one wrapping layer, and at runtime each layer's decoder runs in the correct order. No additional code is needed; concatenation and cross-code composition both fall out of the wrapping mechanism.

\subsection{Implementation}\label{sec:compilation-impl}

The framework provides a \texttt{Compiler} base class that automates the pattern described above. The base class provides \texttt{compile\_kernel} (recursive AST traversal applying handlers), \texttt{wrap\_callbacks} (the wrapping boilerplate from the previous subsection), and a top-level \texttt{compile} method:

\begin{Verbatim}[fontsize=\scriptsize]
def compile(self, program, callbacks=None):
    new_callbacks = self.wrap_callbacks(callbacks or set())
    new_callbacks |= self.compiler_callbacks
    new_kernel = self.compile_kernel(program.kernel)
    return Program(self.target, new_kernel), new_callbacks
\end{Verbatim}

\noindent This method transforms all kernels in the program, wraps all user-provided callbacks, and merges in any callbacks the compiler itself introduces. The output is a new program in the target gate set paired with the adapted callback set. Because each compiler is a standalone object with a \texttt{compile} method, chaining is simply function composition:

\begin{Verbatim}[fontsize=\scriptsize]
p1, c1 = toy_compiler.compile(program, callbacks)
p2, c2 = rep3_bit_compiler.compile(p1, c1)   # bit-flip
p3, c3 = rep3_phase_compiler.compile(p2, c2)  # Shor code
p4, c4 = h2_compiler.compile(p3, c3)
\end{Verbatim}

\noindent Each line produces a self-contained program and callback set that can be passed to the next compiler or directly to an emulator. After the first two QEC passes, the program implements the Shor code with 9 physical qubits per logical qubit. The Toy ISA translation, both repetition codes, Steane, and H2 (trapped-ion) native compilation are all instances of the same \texttt{Compiler} class, differing only in their handlers and decode methods.

The full framework totals approximately 1{,}500 lines of Python, including the compiler base class (71 lines), state-vector emulator, noise models, parser, and Jupyter integration. Individual compiler passes are compact: the Toy-to-Cliffords ISA translator is 30 lines, each repetition-code compiler is approximately 68 lines, and the Steane compiler (the largest) is 231 lines. 
\section{Formal Syntax and Semantics}\label{sec:formal}

We now formalize the IR introduced in Section~\ref{sec:ir}.

\vspace{4pt}
{\small
\begin{tabular}{lrl}
program         &:=& kernel \\
kernel          &:=& $(\textit{qubit\_id}\ :\ \text{instructions}\ :\ \text{callback})$ \\
instructions    &:=& instruction\ instructions  $\mid$ $\varnothing$ \\
instruction     &:=& quantum\_gate $\mid$ kernel \\
quantum\_gate   &:=& \textit{gate\_id}\ \textit{qubit\_id} \ldots \\
callback        &:=& \textit{callback\_id} $\mid$ $\varnothing$ \\
\end{tabular}}

\vspace{4pt}

\noindent Each \textit{gate\_id} denotes a fixed unitary from the program's gate set. Each \textit{callback\_id} denotes a classical function; in our formalism each such function takes a bit string (the current measurement stack) and returns both an optional next kernel and a bit string (the updated measurement stack), i.e.\ a function of type $\text{bit}^* \to \text{kernel option} \times \text{bit}^*$. This is more convenient to formalize than the approach in our implementation, where callbacks mutate the measurement stack via imperative push and pop operations.

We write a kernel as the triple $(q : I : cb)$, corresponding to the concrete syntax \texttt{allocate}~$q$\texttt{:}~$I$~\texttt{measure}~$cb$.

\subsection{Operational Semantics}\label{sec:semantics}

\newcommand{\keval}{\Rightarrow}

\begin{figure*}[t]
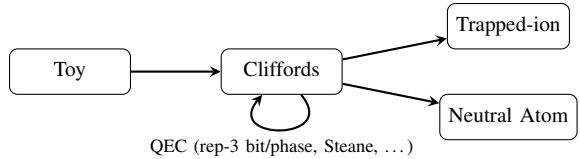

\begin{mathpar}
\inferrule[Start-kernel]
{  }
{ \bigl((q_k:I_k:cb_k)::I,\; cb\bigr)::C,\; S,\; (\ket{\psi^n}, m)
  \;\keval\;
  (I_k, cb_k)::(I, cb)::C,\; S,\; (\ket{\psi^n}\!\otimes\!\ket{0},\; (q_k, n) :: m) }
\and
\inferrule[Quantum-gate]
{  }
{ \bigl(g(q_0,\ldots,q_j)::I,\; cb\bigr)::C,\; S,\; (\ket{\psi}, m)
  \;\keval\;
  (I, cb)::C,\; S,\; (U_g^{(m[q_0],\ldots,m[q_j])}\ket{\psi},\; m) }
\and
\inferrule[End-of-kernel (done)]
{ \|\psi^n\|_{n-1} \mapsto b,\; \ket{\psi'} \\
  m = (q, n-1) :: m' \\
  f_{cb}(b :: S) = (\varnothing,\; S') }
{ ([\,], cb)::C,\; S,\; (\ket{\psi^n}, m)
  \;\keval\;
  C,\; S',\; (\ket{\psi'},\; m') }
\and
\inferrule[End-of-kernel (continue)]
{ \|\psi^n\|_{n-1} \mapsto b,\; \ket{\psi'} \\
  m = (q, n-1) :: m' \\
  f_{cb}(b :: S) = ((q_k:I_k:cb_k),\; S') }
{ ([\,], cb)::C,\; S,\; (\ket{\psi^n}, m)
  \;\keval\;
  (I_k, cb_k)::C,\; S',\; (\ket{\psi'}\!\otimes\!\ket{0},\; (q_k, n-1)::m') }
\end{mathpar}
\vspace{2pt}\hrule
\caption{Small-step operational semantics for qstack programs. Each rule steps a configuration $(C, S, (\ket{\psi}, m))$ where $C$ is the kernel stack, $S$ is the measurement stack, $\ket{\psi}$ is the quantum state vector, and $m$ maps qubit identifiers to positions.}
\label{fig:semantics}
\end{figure*}

We give a small-step operational semantics that captures the interplay between quantum state evolution and classical callbacks. Execution is modeled as a relation on configurations $(C,\; S,\; (\ket{\psi},\, m))$, where:

\begin{itemize}
    \item $C$ is the \emph{kernel stack}: a stack of (instruction-sequence, callback) pairs representing active kernel invocations.
    \item $S$ is the \emph{measurement stack}: a list of classical bits accumulated during execution.
    \item $\ket{\psi}$ is the current quantum state vector.
    \item $m$ maps logical qubit identifiers to positions in the state vector.
\end{itemize}

We write $U_g$ for the unitary associated with gate $g$, and $U_g^{(j_0,\ldots,j_k)}$ for the lifted operator that applies $U_g$ to positions $j_0,\ldots,j_k$ within the full state vector. 

For each callback identifier $cb$, we write $f_{cb}$ for the corresponding classical function of type $\text{bit}^* \to \text{kernel option} \times \text{bit}^*$. The mapping from callback names to functions is provided at runtime and is not part of the program definition; the IR stores only the identifier. In practice, callbacks may also maintain additional internal state and incorporate it into their results; we omit this from the formalization, as it is straightforward for callbacks to manage such state independently.

For measurement, $\|\psi^n\|_{n-1} \mapsto b, \ket{\psi'}$ denotes measuring the last qubit (at position $n{-}1$), yielding classical bit $b$ and post-measurement state $\ket{\psi'}$.

Given a top-level kernel $(q : I : cb)$, the initial configuration is:
$$((I, cb)::[\,],\; [\,],\; (\ket{0},\, (q, 0) :: [\,]))$$
\noindent The kernel's instructions and callback are pushed onto an empty call stack, the measurement stack is empty, the qubit $q$ is allocated at position~0, and the quantum state vector is initialized to $\ket{0}$. A final configuration has the form $([\,],\; S,\; ([\,],\, [\,]))$, where the call stack is empty, $S$ is the program's output (a bit list), and all qubits have been measured and deallocated.

The four inference rules are given in Figure~\ref{fig:semantics}. Which rule applies given the current configuration depends on the form of the instruction at the top of the call stack:

\begin{itemize}
    \item If the next instruction is a nested kernel, \textsc{Start-kernel} applies: it allocates a fresh qubit in $\ket{0}$, extends the qubit mapping, and pushes the nested kernel's instructions onto the call stack.
    \item If the next instruction is a quantum gate, \textsc{Quantum-gate} applies: it applies the corresponding lifted unitary to the state vector. The measurement stack and qubit mapping are unchanged.
    \item If there are no instructions left, the kernel's qubit is measured and deallocated. The measurement bit $b$ is prepended to the measurement stack ($b :: S$), and the callback $f_{cb}$ is invoked with this updated state. 
      \begin{itemize}
        \item If the callback returns no new kernel ($\varnothing$), \textsc{End-of-kernel (done)} applies: control returns to the enclosing kernel. 
        \item If the callback returns a new kernel $(q_k : I_k : cb_k)$, \textsc{End-of-kernel (continue)} applies: the new kernel is initiated immediately, reusing the measured qubit's position in the state vector.
      \end{itemize}
\end{itemize}

\noindent The \textsc{continue} rule is what gives the IR unbounded dynamic computation: repeat-until-success loops, iterative error correction, and conditional fixups are all expressed through callbacks that return new kernels, without any classical control flow in the IR itself. 

\section{Evaluation}\label{sec:evaluation}

We evaluate qstack by compiling a test program through every combination of ISA translation and error correction available in our prototype, then comparing the framework's approach against Qiskit's manual error correction workflow.

\subsection{End-to-End Compilation}\label{sec:fullstack}

We compile a multi-qubit program that allocates qubits, applies gates, and measures one qubit with a conditional callback (\texttt{fix}) that returns a new kernel containing a single gate:

\begin{Verbatim}[fontsize=\scriptsize]
allocate q1:
  allocate q2:
    allocate q3:
      mix q2
      entangle q3
    measure done
  measure fix(q1)
measure done
\end{Verbatim}

\noindent The parameter \texttt{q1} is a constant in the IR: the compiler can see that \texttt{fix} references qubit \texttt{q1}, which is important when compilation remaps qubit identifiers. The \texttt{fix} callback is defined in Python:

\begin{Verbatim}[fontsize=\scriptsize]
def fix(mstack, *, q):
    return Kernel(instructions=[Flip(q)])
\end{Verbatim}

\noindent When invoked, \texttt{fix} returns a kernel that applies a \texttt{Flip} gate to \texttt{q}, which is bound to \texttt{q1} at the call site. The callback is written once against the source gate set; the framework adapts it automatically at every subsequent compilation stage.

Figure~\ref{fig:pipeline} shows the compilation graph. Programs originate in a pedagogical Toy ISA and are first translated to Cliffords, the framework's standard intermediate gate set. From Cliffords, two kinds of passes are available: QEC passes (bit-flip rep-3, phase-flip rep-3, Steane) that encode logical qubits into physical qubits while preserving the Cliffords ISA, and hardware-targeting passes that translate to a device-native gate set. Because QEC passes preserve the ISA, they can be composed freely: applying bit-flip rep-3 followed by phase-flip rep-3 yields the Shor code, applying the same pass twice doubles the code distance, and hardware targeting can follow any number of QEC layers. The graph is also extensible: adding a new hardware backend (for example, a target for superconductor hardware) requires only a single new compiler pass from Cliffords.

\begin{figure}[t]
\centering
\begin{tikzpicture}[
  isa/.style={rectangle, draw, rounded corners=3pt,
              minimum width=1.6cm, minimum height=0.6cm,
              font=\footnotesize\strut},
  planned/.style={isa, draw=gray, text=gray},
  pass/.style={->, thick, >=stealth},
]
  \node[isa] (toy) at (0, 0) {Toy};
  \node[isa] (cliff) at (2.8, 0) {Cliffords};
  \node[isa] (h2) at (5.8, 0.6) {Trapped-ion};
  \node[isa] (na) at (5.8, -0.6) {Neutral Atom};

  \draw[pass] (toy) -- (cliff);
  \draw[pass] (cliff) -- (h2);
  \draw[pass] (cliff) -- (na);

  \draw[pass] (cliff) to[out=-50, in=-130, looseness=4]
    node[below, font=\scriptsize] {QEC (rep-3 bit/phase, Steane, \ldots)} (cliff);
\end{tikzpicture}

\vspace{4pt}
{\footnotesize
\begin{tabular}{@{}cl@{}}
  \textbf{(a)} & Toy $\to$ Cliffords $\to$ Trapped-ion \\
  \textbf{(b)} & Toy $\to$ Cliffords $\to$ rep-3 bit $\to$ Trapped-ion \\
  \textbf{(c)} & Toy $\to$ Cliffords $\to$ Steane $\to$ Neutral Atom \\
  \textbf{(d)} & Toy $\to$ Cliffords $\to$ rep-3 bit $\to$ rep-3 phase $\to$ Trapped-ion (Shor) \\
\end{tabular}}
\caption{Compilation graph and example pipelines. Each edge is a compiler pass. QEC passes (loop) preserve the Cliffords ISA and can be applied repeatedly for concatenation.}
\label{fig:pipeline}
\end{figure}

Table~\ref{tab:stages} summarizes the program state after each compilation step. The key column is the rightmost: the programmer never modifies the \texttt{fix} callback. At the Toy level it returns a kernel containing \texttt{Flip(q)}. After ISA translation, the wrapper compiles this to \texttt{X(q)} (the Clifford equivalent). After bit-flip rep-3 encoding, the wrapper first decodes three physical measurement bits via majority vote, then the inner wrapper compiles the returned kernel, which now operates on three physical qubits per logical qubit. After applying the phase-flip rep-3 pass on top, the program implements the Shor code: each logical qubit is encoded in nine physical qubits, and the two decoding layers run in sequence automatically. Each wrapping layer's decoder runs in the correct order; the callback logic itself is unchanged throughout.

The callback chain makes this layering visible. At the source level, the program references a single callback named \texttt{fix}. After ISA translation, the framework wraps it so that any kernel returned by \texttt{fix} is compiled through the Toy-to-Cliffords pass. After the bit-flip rep-3 pass, a decoder layer is prepended: three physical measurement bits are majority-decoded into one logical bit before the inner chain runs. After the phase-flip rep-3 pass (completing the Shor code), a second decoder is prepended. Each pass adds exactly one wrapping layer; the runtime peels them outside-in on every measurement.

\begin{figure}[t]
\centering
\small
{\footnotesize
\begin{tabular}{@{} l l l @{}}
\toprule
\textbf{Step} & \textbf{ISA} & \textbf{\texttt{fix} kernel} \\
\midrule
Source program     & Toy       & \texttt{Flip(q)} \\
\quad ISA translation    & Cliffords & \texttt{X(q)} \\
\addlinespace
\quad\quad + rep-3 bit         & Cliffords & \texttt{X} on 3 phys.\ qubits \\
\quad\quad\quad + rep-3 phase (Shor) & Cliffords & \texttt{X} on 9 phys.\ qubits \\
\addlinespace
\quad\quad + Steane           & Cliffords & \texttt{X} on 7 phys.\ qubits \\
\addlinespace
\quad\quad\quad $\to$ Trapped-ion  & Trapped-ion & \texttt{U1($\pi$,\,0)} per phys.\ qubit \\
\quad\quad\quad $\to$ Neutral Atom & Neutral Atom & \texttt{SX, SX} per phys.\ qubit \\
\bottomrule
\end{tabular}}
\caption{Program state after each compilation step for a three-qubit test program. The \texttt{fix} callback is written once; each row shows the framework's automatic adaptation. QEC rows at the same indentation level are alternative paths; any combination can precede the hardware-targeting step.}
\label{tab:stages}
\end{figure}

We validated semantic preservation by taking the program after each compilation stage and running it 1{,}000 times on the framework's state-vector emulator without noise, checking that the distribution of final results is the same across all stages. All pipelines produce identical measurement distributions over the logical qubits, confirming that compilation preserves observable behavior across ISA translations, error correction layers, and their compositions.

\subsection{Comparison with Qiskit}\label{sec:qiskit}

To ground our comparison, we implemented a complete Qiskit compilation pipeline mirroring our prototype: custom toy gates, ISA translation passes, and QEC passes (bit-flip rep-3, phase-flip rep-3, and Steane) with syndrome extraction and mid-circuit correction.

Qiskit's natural programming model uses Python to construct circuits, with classical logic residing in the host language at construction time. The typical QEC workflow collects all measurement results, then post-processes them in Python to decode syndromes and determine corrections.

However, Qiskit also exposes circuit-level classical primitives borrowed from OpenQASM~3: typed registers (\texttt{Bool}, \texttt{Uint}), bitwise and comparison operators, \texttt{if\_test}, \texttt{switch}, and variable assignment via \texttt{Store}. These primitives make mid-circuit measurement and correction possible within the circuit itself, and are sufficient for a broad class of decoders: majority vote for repetition codes can be expressed via \texttt{bit\_and}/\texttt{bit\_or}, and lookup-table decoders for Steane or other small codes can be expressed as a \texttt{switch} on a \texttt{Uint} syndrome register.

A compiler can use these primitives to replace each measurement with an inline decoder that maps raw physical results to the decoded logical bit. This approach mirrors qstack's measurement-stack abstraction---in fact, our Qiskit pipeline was directly informed by qstack's design---and enables composition: if each compiler inlines its decoder into every measurement so that decoded results land in the original classical bits, a subsequent compiler can process the resulting circuit, recurse into the control-flow blocks, and produce a correct multi-layer encoding.

The limitation is in automation, not expressivity. Qiskit provides the primitives but not the composition algorithm. Each compiler must be hand-written to handle every gate and classical instruction type, recurse into all \texttt{if\_test} and \texttt{switch} blocks, remap qubit and classical-bit indices correctly across layers, and inline the decoder at exactly the right point. A missing case or incorrect recursion silently produces wrong output.

In qstack, each QEC layer is defined independently and composition follows automatically from the recursive compilation algorithm described in Section~\ref{sec:compilation}. Callbacks serve as that boundary: syndrome extraction and correction live inside a callback; the outer compiler never sees the inner decoder's control flow, so it cannot mangle it. When the outer pass wraps the inner pass's callbacks, it adds a decoding layer that runs before the inner decoder. Each layer's classical logic is composed in the correct order, and returned correction kernels are compiled through the full pipeline.

A second limitation is expressivity. Qiskit's classical operators suffice for codes whose decoders are finite decision procedures---repetition codes, the Steane code, and similar small codes---but not for decoders whose complexity grows with code distance. Minimum-weight perfect matching (MWPM) for surface codes~\cite{dennis2002topological,higgott2022pymatching}, for instance, requires graph algorithms that cannot be serialized into a fixed tree of \texttt{if\_test} and \texttt{switch} nodes. In qstack, callbacks can execute any classical code supported by the runtime environment, so MWPM or any other algorithm can be invoked directly.

\subsection{OpenQASM~3}\label{sec:openqasm}

Qiskit circuits compile down to OpenQASM~3~\cite{cross2022openqasm}, so the automation and composition considerations from Qiskit apply equally at the language level. The control-flow constructs available in Qiskit (\texttt{if\_test}, \texttt{switch}, typed classical registers) correspond directly to OpenQASM~3 primitives, and the \texttt{def} subroutine mechanism provides a callable boundary around mixed classical-quantum logic that a compiler could use to encapsulate decoders.

The other capability OpenQASM~3 adds beyond Qiskit is \texttt{extern}: a classical function resolved at link time that provides an escape hatch to arbitrary computation. A surface-code decoder based on MWPM could be invoked via \texttt{extern} to compute syndrome-to-correction mappings at runtime, closing the expressivity gap identified above. However, \texttt{extern} is restricted to classical types: it can return an integer syndrome label, but not quantum instructions. Mapping the result back to corrections still requires a \texttt{switch} over pre-enumerated branches, which are static circuit elements.

What remains missing is the compiler algorithm. OpenQASM~3 is an intermediate representation, not a compiler framework. The algorithm that generates composed decoders, remaps qubits and syndrome registers across layers, and chains decoder calls in the correct order is not part of the language specification. This algorithm is precisely what qstack contributes. In qstack, callbacks unify both aspects: they serve as composable subroutine boundaries (like \texttt{def}) while also supporting arbitrary classical logic including algorithms like MWPM (like \texttt{extern}), and the result is a kernel of quantum instructions that the framework compiles through the full pipeline.

These observations suggest a path forward: qstack's compilation model could be realized on OpenQASM~3 using \texttt{def} as the wrapping boundary and \texttt{extern} for decoder expressivity, bringing composable QEC compilation to a widely adopted quantum assembly language.

\section{Related Work}\label{sec:related}

\subsection{Quantum Programming Frameworks}

The dominant approach to quantum programming embeds circuit construction in a host language. Qiskit~\cite{Qiskit}, Cirq~\cite{cirq}, and PyQuil~\cite{pyquil} use Python as the host; Quipper~\cite{green2013quipper} uses Haskell. This design is flexible but blurs the boundary between compile-time circuit construction and run-time execution: classical control logic resides in the host language rather than in a representation the compiler can analyze or transform.

Q\#~\cite{qsharp}, Silq~\cite{silq}, and Guppy~\cite{guppy} are standalone quantum languages with built-in classical constructs, but their compilers must reason about interleaved classical and quantum code. OpenQASM~3~\cite{cross2022openqasm} and Quil~\cite{smith2016practical} extend assembly-level circuit descriptions with classical types and control flow for hybrid programs. Adding classical expressions to a quantum language effectively creates two languages that must coexist in a single compiler. OpenQASM illustrates this: it began with classical-bit-controlled operations sufficient for teleportation-style fixups, but has been steadily extended to support integers, booleans, and loops as more algorithms demanded richer control. Yet real-time decoders for quantum error correction may require arbitrarily complex classical logic---table lookups, iterative algorithms, or calls to external libraries---that exceeds what any fixed set of in-language primitives can reasonably provide, while the quantum compiler must understand and correctly transform every classical construct it encounters.

Qunity~\cite{voichick2023qunity} avoids this by eliminating classical computation entirely, but without classical feedback it cannot express the real-time syndrome decoding and conditional correction that quantum error correction requires.

qstack takes a different approach: classical logic is entirely opaque, accessed only through callbacks. The compiler never reasons about classical code; it only ensures that callbacks receive correctly decoded measurement results and that returned kernels are compiled through the full pipeline. Classical optimizations (such as branch elimination) are not lost; they are delegated to the classical language's own compiler, which can apply them independently. The result is a is fully compositional compilation framework.

\subsection{QEC Tools}

Stim~\cite{gidney2021stim} is a high-performance stabilizer circuit simulator that handles tens of thousands of qubits through Pauli-frame simulation. The qiskit-qec library~\cite{qiskit-qec} provides QEC utilities within the Qiskit ecosystem. Lattice surgery compilers~\cite{Litinski2019gameofsurfacecodes} translate logical operations into fault-tolerant layouts for topological codes.

These tools target specific QEC tasks (simulation, syndrome decoding, layout). qstack instead integrates error correction as ordinary compiler passes within a general compilation pipeline. The callback wrapping mechanism allows QEC passes to compose with ISA translation and with each other (for code concatenation) without specialized infrastructure.

\section{Conclusion}\label{sec:conclusion}

We have presented qstack, a compiler framework for quantum programs built around two complementary ideas. First, a minimal IR based on the kernel model (allocate/compute/measure/callback) provides a uniform representation at every stage of the compilation pipeline: the same format describes programs in a high-level pedagogical gate set, in Clifford gates, in error-corrected encodings, and in hardware-native instructions. Second, callback wrapping enables compositional compilation over this IR: each compiler pass automatically adapts classical callbacks so that measurement results are correctly decoded and any dynamically generated kernel is compiled through the full pipeline. Together, the uniform IR and the wrapping mechanism allow ISA translation and quantum error correction passes to be freely composed, including concatenation of error-correcting codes, without manual intervention.

Our prototype demonstrates compilation from a high-level gate set through Clifford gates to trapped-ion native operations, with bit-flip and phase-flip repetition codes, the Shor code (obtained by composing the two), and the Steane code as QEC layers that can be inserted at any point. Because the IR is the same at every stage, adding a new hardware target or a new QEC code requires implementing only gate handlers and a decode method; the wrapping mechanism and compositional infrastructure are provided by the framework.

Several directions remain open. First, the compositional structure of qstack is well-suited to formal verification: each pass can be verified independently, and a one-time proof of the wrapping mechanism could yield end-to-end correctness guarantees. Second, scaling to more complex codes (surface codes with adaptive decoders, color codes) will test whether the callback abstraction remains practical when decoding requires iterative classical computation. Third, integrating with real hardware backends will require bridging qstack's abstract qubit model with device-specific constraints such as connectivity, timing, and calibration data. The combination of a uniform IR with compositional callback wrapping provides a solid foundation for these extensions.

\bibliographystyle{IEEEtran}
\bibliography{references}

\end{document}